\documentclass[aps,prd,twocolumn,10pt,showpacs]{revtex4-1}
\usepackage[mathcal]{euscript}
\usepackage[latin1]{inputenc}
\usepackage{latexsym}
\usepackage{amsmath}
\usepackage{hyperref}
\usepackage{amsmath}    
\usepackage{graphicx}   
\usepackage{verbatim}   
\usepackage{color}      
\usepackage{subfigure}  
\usepackage{bm}
\usepackage{amssymb}
\usepackage{bbm}
\usepackage{cancel}

\begin{document}
\setcounter{topnumber}{1}

\title{Two formalisms, one renormalized stress-energy tensor}
\author{C. Barcel\'{o}}
\affiliation{Instituto de Astrof\'{i}sica de Andaluc\'{i}a, CSIC, Glorieta de la Astronom\'{i}a, 18008 Granada, Spain}
\author{R. Carballo}
\affiliation{Instituto de Astrof\'{i}sica de Andaluc\'{i}a, CSIC, Glorieta de la Astronom\'{i}a, 18008 Granada, Spain}
\author{L. J. Garay}
\affiliation{Departamento de F\'{i}sica Te\'{o}rica II, Universidad Complutense de Madrid, 28040 Madrid, Spain}
\affiliation{Instituto de Estructura de la Materia, CSIC, Serrano 121, 28006 Madrid, Spain}

\begin{abstract}

We explicitly compare the structure of the renormalized stress-energy tensor (RSET) of a massless scalar field in a (1+1) curved spacetime as obtained   by two different strategies: normal-mode construction of the field operator and one-loop effective action. We pay special attention to where and how it is encoded the information related to the choice of vacuum state in both formalisms. By establishing a clear translation map between both procedures, we show that these two potentially different RSET are actually equal, when using vacuum-state choices related by this map. One specific aim of the analysis is to facilitate the comparison of results regarding semiclassical effects in gravitational collapse as obtained within these different formalisms.

\end{abstract}

\maketitle


\section{Introduction}

One of the central problems of Quantum Field Theory in Curved Spacetimes is the calculation of the renormalized stress-energy tensor (RSET) of the matter fields. This RSET is then the starting point of any calculation in Semiclassical Gravity: Spacetime is maintained as a classical entity while the classical stress-energy tensor of the matter sources is complemented with the RSET of the quantum fields in a suitable vacuum state.  

In general, the calculation of the RSET is complicated and cannot be done analytically. However, if the spacetime manifold has some degree of symmetry 
the reduced RSET can sometimes be obtained explicitly. Here we are interested in spacetimes which can be reduced by symmetry to become effectively (1+1)-dimensional. More specifically we are going to consider a massless scalar field theory adapted to this (1+1) symmetry reduction. For instance, we shall have in mind four-dimensional spherically symmetric stellar configurations, including black-hole spacetimes. 

Here, by spherically symmetric reduction we understand the quantization of the classically reduced system, which implies not only the assumption of a spherically symmetric classical background, but also neglecting the effect of any non-spherically symmetric fluctuations. This quantization is a toy-model approximation to the full quantization.

 A spherical-symmetry reduction makes the scalar field equation to become (1+1)-dimensional. After neglecting the effect of a potential term, or what is equivalent, after neglecting backscattering processes, the field equation ends up corresponding to a massless Klein-Gordon equation in a curved spacetime of (1+1) dimensions. This equation is conformally invariant, which is the crucial ingredient for the analytical tractability of the reduced problem. The exact and analytic RSET calculated in this way is, therefore, a toy-model  approximation to the four-dimensional spherically-symmetric RSET. However, this RSET provides crucial insights about the physics at work in the actual (3+1) setting 
For example, one can perfectly describe the Hawking radiation process except for the grey-body factors~\cite{Page:1976df}.

In the literature there exist two main methods to obtain an analytical expression for the RSET of a massless scalar field in (1+1) dimensions: One is the construction of the stress-energy tensor operator directly from a normal-mode expansion of the field operator~\cite{dfu1,dfu}; and the other is based on an effective-action calculation~\cite{Mottola:2006ew,mukh}. In principle,  these two   different methods  could lead to different expressions for the RSET, although both formalisms should be equivalent due to Wald's theorem \cite{Wald:1977up,full}.

 
  However, the problem is that at first sight the two expressions obtained are difficult even to compare because they use different fiduciary structures. In this paper we work out in detail the relationship between them, with the aim of providing a common framework to compare the results obtained with both methods when studying specific problems. In particular, we compare how and where it is encoded the information related to the vacuum-state choice in both formalisms. Establishing this translation could be useful, for instance, in studying the effects of back reaction in gravitational collapse. In this context the expressions of the RSET derived following these two different  formalisms  are being used by different authors to study the possibility of avoiding black-hole formation due to quantum effects~\cite{Barcelo:2007yk,Mottola:2006ew}. 

The organization of this article is as follows. First, in section~\ref{Sec:setup}, we will recall these two methods for the calculation of the RSET. Then, in section~\ref{Sec:rel1} we will show that the two expressions found using  normal-mode  and effective-action techniques are actually equal. We will proceed in next section~\ref{Sec:rel2} to describe the connection between the vacuum-state choice in both formalisms. To complete these formal developments we present in section \ref{Sec:deg} a discussion about degeneracy in the vacuum selection and its implications. In section \ref{Sec:mott} we sketch an application of the translation map between the two formalisms in the particular case of a simplified model of stellar collapse, finishing then the article with a brief summary section.

\section{Setup of the problem \label{Sec:setup}}

As already mentioned, in this paper we shall deal with a simple scalar field $\phi$ defined over a globally-hyperbolic (1+1)-dimensional spacetime manifold $\mathcal{M}$, obeying the massless Klein-Gordon equation
\begin{equation}
\square\phi=0~.
\label{eq:eqmotion}
\end{equation}
(We will not distinguish between $\phi$ and its quantum counterpart $\hat{\phi}$ unless we find it confusing.) This equation is invariant under conformal transformations of the metric that leave the field unaffected (conformal transformation of weight zero). 
This equation can be derived from an action 
\begin{equation}
S[\phi,g]=-\frac{1}{2}\int \text{d}^2x\sqrt{-g}\partial_\mu\phi\partial^\mu\phi~,
\label{eq:action}
\end{equation}
which is also conformally invariant.

In (1+1)-dimensions all metrics are conformal to the flat metric. For instance, by using null coordinates $(u,v)$ we can write the line element as
\begin{equation}
ds^2=-C(u,v)\text{d}u\text{d}v~.
\label{eq:linel}
\end{equation}
In these coordinates the equation of motion (\ref{eq:eqmotion}) reads
\begin{equation}
\frac{\partial^2\phi}{\partial u\partial v}=0~.
\label{eq:eqmotionnull}
\end{equation}
A set of positive-norm (in the Klein-Gordon product (\ref{eq:kgprod}) below) mode solutions is
\begin{equation}
\phi^u_\omega:=\frac{1}{\sqrt{4\pi\omega}}\mbox{e}^{-i\omega u}~,\quad
\phi^v_\omega:=\frac{1}{\sqrt{4\pi\omega}}\mbox{e}^{-i\omega v}~,\quad \omega\in\mathbb{R}^+ ~.
\label{eq:orto}
\end{equation}

These solutions are not in the space $\mathfrak{S}_0$ of real solutions of (\ref{eq:eqmotion}), but in the complexified space $\mathfrak{S}^\mathbb{C}_0$. In this complexified space of solutions we can define the pseudo-inner product
\begin{equation}
(\phi_1,\phi_2)_{\mbox{\tiny KG}}:=
i\int_{\Sigma}\text{d}\Sigma^\mu(\phi_1^*\partial_\mu\phi_2-\phi_2\partial_\mu\phi_1^*)~,
\label{eq:kgprod}
\end{equation}
where $\Sigma$ is a Cauchy surface. This product is not positive definite: if an element $\phi\in\mathfrak{S}^\mathbb{C}_0$ has   positive norm, then $\phi^*\in\mathfrak{S}^\mathbb{C}_0$ has  negative norm. By using a null Cauchy surface in the integral above it is direct to see that the mode solutions (\ref{eq:orto}) verify
\begin{align}
(\phi^a_\omega,\phi^b_{\omega'})&=\delta(\omega-\omega')\delta_{ab}~,\nonumber\\
({\phi^a}^*_\omega,{\phi^b}^*_{\omega'})&=-\delta(\omega-\omega')\delta_{ab}~,\nonumber\\
({\phi^a}_\omega,{\phi^b}_{\omega'}^*)&=0~,
\end{align}
where both labels $a,b$ can take the values $u,v$. The general real solution of (\ref{eq:eqmotion}) (i.e. an element of $\mathfrak{S}_0$) can be then written as
\begin{eqnarray}
\hspace{-4mm}\phi=\int_{0}^{\infty}\hspace{-3mm}\text{d}\omega\left(a^u_\omega\phi^u_\omega+a^v_\omega\phi^v_\omega+\mbox{h.c.}\right)~.
\label{eq:confexp}
\end{eqnarray}

As it is well known, there are many alternative ways in which one can choose to expand arbitrary real solutions of $\mathfrak{S}_0$. In selecting a particular expansion, the choice of a particular time function $t$ plays a fundamental role. In fact, the orthonormal modes just described have positive frequency with respect to the timelike vector field $\partial_t:=\partial_u+\partial_v$,
\begin{equation}
i\partial_t\phi^a_\omega=\omega\phi^a_\omega~.
\label{eq:posfreq}
\end{equation}
Actually, there are as many sets of mode solutions of this type (\ref{eq:orto}) as reparametrizations of the null coordinates
\begin{equation}
u\longrightarrow U(u)~,\ \ \ \ \ \ \ \ \ \ v\longrightarrow V(v)~,
\label{eq:rep}
\end{equation}
none of them being preferred to any other. These reparametrizations induce a change in the conformal factor of the metric (\ref{eq:linel}) of the form
\begin{equation}
C\longrightarrow C\frac{\text{d}u}{\text{d}U}\frac{\text{d}v}{\text{d}V}~.
\label{eq:confchange}
\end{equation}
Notice that one needs $\text{d}U/\text{d}u>0$ and $\text{d}V/\text{d}v>0$ so that these transformations  preserve the time orientation of the spacetime.

\subsection{RSET from normal-mode expansion}

The promotion of the creation and annihilation variables to operators permits us to define the vacuum state $|0\rangle$ of the field. The rest of the states of the Fock space will be obtained straightforwardly by the action of the creation operators on this vacuum. We are interested in the vacuum expectation value of the stress-energy tensor:
\begin{equation}
\langle0|\hat{T}_{\mu\nu}(x)|0\rangle.
\label{eq:rsetunr}
\end{equation}
Here $\hat{T}_{\mu\nu}(x)$ is the result of turning the creation and annihilation variables into operators in the functional expression of the classical stress-energy tensor $T_{\mu\nu}(x)$. However, without further considerations the expression (\ref{eq:rsetunr}) is divergent. Thus, to extract a meaningful finite result one has to carry out a renormalization procedure.

One well known renormalization procedure is the point-splitting method~\cite{dfu}, which uses the formal identity
\begin{equation}
\langle0|\hat{T}_{\mu\nu}(x)|0\rangle=\lim_{x'\rightarrow x}\mathcal{D}_{\mu\nu}(x,x')G^{(1)}(x,x')~,
\label{eq:ren1}
\end{equation}
where $G^{(1)}$ is the Green function
\begin{equation}
G^{(1)}(x,x')=\langle0|\{\hat{\phi}(x),\hat{\phi}(x')\}|0\rangle~,
\end{equation}
and $\mathcal{D}_{\mu\nu}(x,x')$ is a suitable differential operator. Before taking the limit in (\ref{eq:ren1}), one has a well defined quantity; it is in the coincidence limit $x\rightarrow x'$ when the divergent behavior appears. To eliminate this divergence, one subtracts a function $S(x,x')$ with the same divergent behavior in the limit $x\rightarrow x'$.

A state is said to be Hadamard if the singular behavior of its Green function is the natural generalization to curved spacetime of its singular structure in Minkowski spacetime. In this case the function $S(x,x')$ takes the form of a Hadamard distribution~\cite{Wald:2009uh}. These states are the ones that are typically considered as physical and are the ones we shall deal with here.

The function $G^{(1)}(x,x')$ can be constructed using the modes (\ref{eq:orto}), and it defines a Hadamard state. After some computations, the vacuum expectation value of the renormalized stress-energy tensor takes the form~\cite{dfu,birrell-davies}
\begin{eqnarray}24\pi\langle0|\hat{T}_{uu}|0\rangle&&=\frac{1}{C}\frac{\partial^2C}{\partial u^2}-\frac{3}{2C^2}\left(\frac{\partial C}{\partial u}\right)^2~,\nonumber\\
24\pi\langle0|\hat{T}_{vv}|0\rangle&&=\frac{1}{C}\frac{\partial^2C}{\partial v^2}-\frac{3}{2C^2}\left(\frac{\partial C}{\partial v}\right)^2~,
\nonumber\\
24\pi\langle0|\hat{T}_{uv}|0\rangle&&=24\pi\langle0|\hat{T}_{vu}|0\rangle=-\frac{R}{4}C~,
\label{eq:rset}\end{eqnarray}
where $C=C(u,v)$ is the conformal factor in the line element (\ref{eq:linel}). 

Picking a different set of modes, replacing $u\rightarrow U(u)$ and $v\rightarrow V(v)$ in (\ref{eq:orto}) or,  in other words, selecting a different time function $t'$, defines a different quantization of the classical theory. In fact, these quantizations can be even unitarily inequivalent, in the sense that there may not exist an unitary map between their corresponding Fock spaces. The new RSET in the new $(U,V)$ coordinate system is the result of replacing the conformal factor in the right side of (\ref{eq:rset}) using the rule (\ref{eq:confchange}).

In the literature, the vacuum states associated with the class of modes (\ref{eq:orto}) are usually called conformal (for reasons that we will see later, which do not imply the invariance of these vacua under {\em general} conformal transformations). The set of conformal vacua is by construction in direct correspondence with the set of reparametrizations of null coordinates. As we will recall, this correspondence is not strictly one-to-one, as there exists a finite set of reparametrizations which do not change the vacuum state.

A priori other non-conformal vacuum states could be defined in the two-dimensional theory. For them the renormalized stress-energy tensor would have to be calculated by  other methods.

\subsection{RSET from effective action}

A different starting point to obtain the RSET is to focus on the dynamics of a classical gravitational field $g_{\mu\nu}$ when it is coupled to the quantum field $\phi$. The classical action of the system is a sum of two terms,
\begin{equation}
S_0[g]+S[\phi,g]~.
\label{eq:classact}
\end{equation}
Here, the action $S_0[g]$ describes the free dynamics of the gravitational degrees of freedom. In the standard case it corresponds to the Einstein-Hilbert action (although for renormalization procedures to make sense it must contain higher-derivative terms; this point is not relevant in what follows). The action for the scalar field $S[\phi,g]$ is given by (\ref{eq:action}). If one treats the spacetime metric as a classical field and considers $\phi$ as a quantum fluctuating field
it is expected that these fluctuations exert some influence over the classical dynamics dictated by (\ref{eq:classact}). The effective action which accounts for this modification of the classical dynamics is found by integrating out the matter degrees of freedom in the path integral formulation. At one-loop, the total action reads~\cite{mukh,barv}
\begin{equation}
S_0[g]+\Gamma[g]~,
\label{eq:totact}
\end{equation}
with
\begin{equation}
\mbox{e}^{i\Gamma[g]}=\int\mathcal{D}\phi\,\mbox{e}^{iS[\phi,g]}~.
\label{eq:effact}
\end{equation}
%
The effective equations of motion for the metric which arise from (\ref{eq:totact}) are
\begin{equation}
\frac{\delta S_0[g]}{\delta g^{\mu\nu}(x)}+\frac{\delta \Gamma[g]}{\delta g^{\mu\nu}(x)}=0~.
\end{equation}
The first term of this equation describes the classical dynamics of the gravitational field whereas the second one
%
represents the quantum corrections and plays the role of an effective stress-energy tensor. We will denote it $\sqrt{-g}\langle \hat{T}_{\mu\nu}(x)\rangle/2$ in what follows.

At this stage  expression~(\ref{eq:effact}) for the one-loop effective action $\Gamma[g]$ (and hence for $\langle\hat{T}_{\mu\nu}\rangle$) is only formal. Making sense of this expression requires the prescription of a specific calculation procedure (see for example~\cite{mukh}). 
First, one calculates the Euclidean effective action by selecting some boundary conditions for the expansion of the relevant Euclidean differential operator in eigenfunctions and by choosing an integration measure ${\cal D}\phi$ adapted to this expansion. 
The resulting expression involves the product of eigenvalues of this Euclidean boundary problem, but is still formally divergent. Thus, one has to use a renormalization procedure to obtain a finite quantity with physical significance. For example, one can use a zeta-function regularization \cite{mukh,Hawking:1976ja}. Finally, one has to extend this expression to the Lorentzian sector.

In two dimensions, the procedure sketched above  leads to
\begin{equation}
\Gamma[g]=\frac{1}{96\pi}\int \text{d}^2x\sqrt{-g(x)}\,R(x)(\square^{-1}R)(x)~.
\label{eq:effact1}
\end{equation}
%
 In this formulation, the vacuum-state choice for the field shows up in the 
extension to the Lorentzian sector, owing to the non-uniqueness of the symbol $\square^{-1}$ (the inversion of an operator with a non-vanishing kernel).
To find the RSET in this vacuum state one only needs to functionally differentiate (\ref{eq:effact1}), resulting in \cite{mukh}
\begin{eqnarray}
24\pi\langle\hat{T}_{\mu\nu}\rangle&&=Rg_{\mu\nu}-\nabla_\mu\nabla_\nu(\square^{-1}R)
\nonumber\\
&&\hspace{-3mm}+\frac{1}{2}\nabla_\mu(\square^{-1}R)\nabla_\nu(\square^{-1}R)
\nonumber\\
&&\hspace{-3mm}-\frac{1}{4}g_{\mu\nu}\nabla^\alpha(\square^{-1}R)\nabla_\alpha(\square^{-1}R)~.
\label{eq:nonlocal}
\end{eqnarray}
This expression is nonlocal, but it can be converted into a local form by introducing a scalar field
\begin{equation}
\varphi(x):=-(\square^{-1} R)(x)=-\int \text{d}^2x'\,G_F(x,x')R(x')~,
\end{equation}
where $G_F$ is a Feynman Green function which carries the information about the choice of vacuum state. Then, expression (\ref{eq:nonlocal}) becomes
\begin{eqnarray}24\pi\langle0_\varphi|\hat{T}_{\mu\nu}|0_\varphi\rangle=\nabla_\mu\nabla_\nu\varphi-g_{\mu\nu}\square\varphi+\nonumber\\
+\frac{1}{2}\nabla_\mu\varphi\nabla_\nu\varphi-\frac{1}{4}g_{\mu\nu}\nabla_\alpha\varphi\nabla^\alpha\varphi~.
\label{eq:rsetmott}
\end{eqnarray}
The subindex $\varphi$ used in $|0_\varphi\rangle$ remarks that selecting a specific solution of the inhomogeneous equation (based on some physical requirements)
\begin{equation}
\square\varphi=-R
\label{eq:varphieq}
\end{equation}
is equivalent to selecting a  specific Feynman Green function and its corresponding  vacuum state whose RSET is (\ref{eq:rsetmott}). The set of real solutions of this equation will be denoted by $\mathfrak{S}_R$.

\section{Comparison between the two formalisms}

We have sketched the two most popular methods to obtain the RSET. Wald's theorem \cite{Wald:1977up,full} asserts that, under natural physical conditions, any two different   procedures to calculate the RSET have to lead to equivalent expressions up to a covariantly conserved tensor locally constructed from the curvature. Indeed, these two approaches lead to RSET expressions which verify the conditions of the theorem. However, at first sight  the comparison of the two expressions is not direct. This is due to the fact that they seem to use different fiduciary structures. In the following we are going to establish a clear translation dictionary between them to enable the cross-fertilization of both approaches.


\subsection{Relationship between the RSET in both approaches \label{Sec:rel1}}

Let us begin by comparing the two expressions for the RSET (\ref{eq:rset}) and (\ref{eq:rsetmott}) of a certain fixed vacuum state. To do this, we will first build  a covariant expression for the RSET which reduces to (\ref{eq:rset}) when we evaluate it in the $(u,v)$ coordinates. A possibility for doing so is to look for a geometric quantity related to the vacuum selection. In this regard it is suggestive to realize that the norm of the vector field $\xi:=\partial_t=\partial_u+\partial_v$, which enters into the positive frequency condition (\ref{eq:posfreq}), is
\begin{equation}
g(\xi,\xi)=-|\xi|^2=-C(u,v)~,
\label{eq:killmod}
\end{equation}
where $C(u,v)$ is precisely the conformal factor which appears in the line element (\ref{eq:linel}). In fact, this identity permits us to write the RSET~(\ref{eq:rset}) in a covariant form:
\begin{eqnarray}
\langle0_{\xi}|\hat{T}_{\mu\nu}|0_{\xi}\rangle=\frac{R}{48\pi}g_{\mu\nu}+\frac{1}{48\pi}\left(A^\xi_{\mu\nu}-g_{\mu\nu}A^\xi/2\right),\nonumber\\
A^\xi_{\mu\nu}:=\frac{4}{|\xi|}\nabla_\mu\nabla_\nu|\xi|=\frac{2}{|\xi|}\nabla_\mu\left(\frac{1}{|\xi|}\nabla_\nu|\xi|^2\right)~.
\label{eq:covrset}
\end{eqnarray}
In these expressions we have denoted by $|0_\xi\rangle$ the vacuum state associated with the positive frequency condition (\ref{eq:posfreq}) with respect to $\xi$. To show that (\ref{eq:covrset}) reproduces (\ref{eq:rset}) it is sufficient to particularize to $(u,v)$ coordinates, using (\ref{eq:killmod}) and the nonzero Christoffel symbols of the metric~(\ref{eq:linel}):
\begin{equation}
\Gamma^{u}_{uu}=\frac{1}{C}\frac{\partial C}{\partial u}~,\ \ \ \ \ \ \ \ \ \ \Gamma^{v}_{vv}=\frac{1}{C}\frac{\partial C}{\partial v}~.
\label{eq:christ}
\end{equation}

We have to compare two covariant expressions for the renormalized energy-momentum tensor, (\ref{eq:rsetmott}) and (\ref{eq:covrset}), which, taking into account Wald's theorem, suggests a relationship between $|\xi|$ and $\varphi$. In two-dimensional spacetimes using null coordinates $(u,v)$, we have the identity
\begin{equation}
R=-\square\log C(u,v)~.
\end{equation}
Thus, a relationship which guarantees that Eq. (\ref{eq:varphieq}) is fulfilled is
\begin{equation}
\varphi=\log|\xi|^2~.
\label{eq:relation1}
\end{equation}
Then, it is easy to see, using
\begin{align}
\frac{1}{|\xi|}\nabla_\mu\nabla_\nu |\xi|
&=\frac{1}{2}\nabla_\mu\nabla_\nu\varphi+\frac{1}{4}\nabla_\mu\varphi\nabla_\nu\varphi~
\end{align}
and the equation of motion (\ref{eq:varphieq}), that
\begin{align}
\langle0_{\xi}|\hat{T}_{\mu\nu}|0_\xi\rangle&=\frac{1}{2}Rg_{\mu\nu}+\nabla_\mu\nabla_\nu\varphi+\frac{1}{2}\nabla_\mu\varphi\nabla_\nu\varphi
\nonumber\\
&-\frac{1}{2}g_{\mu\nu}\left(\square\varphi+\frac{1}{2}\nabla_\alpha\varphi\nabla^\alpha\varphi\right)
\nonumber\\
& =\langle0_\varphi|\hat{T}_{\mu\nu}|0_\varphi\rangle.
\end{align}
Therefore the two expressions for the RSET are {\em actually equal}. Moreover, as we will explain in detail in the next section, this faithful equivalence provides through (\ref{eq:relation1}) a map between the possible choices of vacuum state in both formalisms.

\subsection{Relationship between vacuum states in both approaches \label{Sec:rel2}}

The vacuum states $|0_\varphi\rangle$ are in correspondence with the space of solutions $\mathfrak{S}_R$ of the inhomogeneous equation~(\ref{eq:varphieq}). To define an application between the two sets of vacua, $\{|0_\varphi\rangle\}$ and $\{|0_\xi\rangle\}$, we must characterize the set of vector fields which appear in the positive-frequency condition (\ref{eq:posfreq}).

The vector field $\partial_t=\partial_u+\partial_v$ which appears in (\ref{eq:posfreq}) is a conformal Killing vector field, as it can be shown by direct computation (this is the reason for the adjective ``conformal'' attached in the literature to the corresponding vacuum state). In fact, all the conformal Killing vector fields over $\mathcal{M}$ can be written in this form in appropriate null coordinates as is shown in what follows. If $\xi$ is a conformal Killing vector field, it verifies the conformal Killing equations
\begin{equation}
\nabla_\mu\xi_\nu+\nabla_\nu\xi_\mu
\propto g_{\mu\nu}
\label{eq:confk}
\end{equation}
%
As in $(u,v)$ coordinates the metric is (\ref{eq:linel}) and the nonzero Christoffel symbols are (\ref{eq:christ}), given an arbitrary vector field
\begin{equation}
\xi=a(u,v)\partial_{u}+b(u,v)\partial_{v}~,
\end{equation}
equations (\ref{eq:confk}) become
\begin{equation}
\frac{\partial b(u,v)}{\partial u}=0~,\qquad \frac{\partial a(u,v)}{\partial v}=0~,
\end{equation}
(the equation for $\mu=u$, $\nu=v$ is automatically verified). Such a conformal Killing vector field can then be written as
\begin{equation}
\xi=a(u)\partial_{u}+b(v)\partial_{v}=\partial_{U}+\partial_{V}~,
\end{equation}
where we have defined the null coordinates $(U,V)$ by
\begin{equation}
\frac{\text{d}U}{\text{d}u}=\frac{1}{a(u)}~,\qquad \frac{\text{d}V}{\text{d}v}=\frac{1}{b(v)}~.
\end{equation}
Assuming that the time orientation of the metric manifold is such that the null coordinates $(u,v)$ are running towards the future, the preservation of this time orientation is equivalent to require the inequalities $\text{d}U/\text{d}u>0$ and $\text{d}V/\text{d}v>0$ to hold. Then both $t=(u+v)/2$ and $t'=(U+V)/2$ are forward-running time functions. 

Notice that the spacetime manifold might not admit a global timelike conformal Killing vector field. Moreover, even if the manifold admits a global timelike Killing vector field, we could still define vacuum states by attending to their properties in a local patch of the manifold, even if their associated conformal vector fields $\xi$ were not globally timelike. For instance, the vector field could become null in some specific places. These choices will give rise to vacuum states with singular behavior at these places. Although unphysical, these vacuum states could be of interest as the limits of families of well defined vacua. This singular behavior is precisely what happens to the Boulware state in the event horizon of a static black hole \cite{Boulware:1974dm}. Hereafter, we will assume by default that all our assertions apply at least to a local patch of the spacetime manifold.

Let us denote by $\mathfrak{X}_{\mbox{\tiny CK}}^+(\mathcal{M})$ the set of future-oriented timelike conformal Killing vectors fields. We have shown that this set is in direct correspondence with the set of possible future-oriented null coordinates and, by construction, with the set of mode expansions of the type~(\ref{eq:confexp}), each corresponding to a different choice of conformal vacuum. More explicitly, to select one of these mode expansions one imposes the requisite of positive frequency of the modes (\ref{eq:posfreq}) with respect to some $\xi\in\mathfrak{X}_{\mbox{\tiny CK}}^+(\mathcal{M})$.
Technically, this choice can be seen to correspond to the selection of a complex structure  over the symplectic manifold of real solutions $\mathfrak{S}_0$ of the scalar field equation  \cite{Corichi:2002qd}. Let us define the complex structure as
\begin{equation}
J_\xi:=-\frac{1}{|\mathcal{L}_\xi|}\mathcal{L}_\xi~,
\label{eq:Jdef}
\end{equation}
where $\mathcal{L}_\xi$ represents the Lie derivative along $\xi$, whose action (extended to complex solutions) is
\begin{equation}
J_\xi\phi_\omega=i\phi_\omega~.
\end{equation}
Using  expansion (\ref{eq:confexp}) it can be shown that (\ref{eq:Jdef}) is a complex structure over $\mathfrak{S}_0$, defining a quantum theory and a vacuum state $|0_\xi\rangle$. In fact, we can recognize the complex structure (\ref{eq:Jdef}) as the natural one when the spacetime admits a timelike Killing vector field $\xi$ \cite{Corichi:2002qd,Ashtekar:1975zn}. 

The set of vacua $\{|0_\xi\rangle\}$ is in direct correspondence with $\mathfrak{X}_{\mbox{\tiny CK}}^+(\mathcal{M})$, so it is parametrized by two real functions of one variable given by the reparametrizations of the null coordinates (\ref{eq:rep}). As we will see, except for a
finite-dimensional group of reparametrizations (see next section) all these complex structures are different.

Within this setting we can promote (\ref{eq:relation1}) to a map between the set of all future-oriented timelike conformal Killing vectors and the set of solutions of Eq.~(\ref{eq:varphieq})
\begin{equation}
\gamma:\ \mathfrak{X}_{\mbox{\tiny CK}}^+(\mathcal{M})\longrightarrow\mathfrak{S}_R~,
\label{eq:gamma}
\end{equation}
given by
\begin{equation}
\gamma(\xi)=\varphi:=\log|\xi|^2.
\end{equation}
This map (\ref{eq:gamma}) is bijective. The inverse map is the following. Picking a set of null coordinates $(u,v)$ the general solution of (\ref{eq:varphieq}) can be written as
\begin{equation}
\varphi(u,v)=\log C(u,v)+f(u)+g(v)~,
\end{equation}
with $f$, $g$ being real-valued functions, solutions of the homogeneous part of the equation (\ref{eq:varphieq}). The corresponding conformal Killing vector field is then
\begin{equation}
\xi=\gamma^{-1}(\varphi)=a(u)\partial_u+b(v)\partial_v~,
\end{equation}
with
\begin{equation}
a(u)=\exp[f(u)]~,\qquad
b(v)=\exp[g(v)]~.
\end{equation}
To summarize, the map (\ref{eq:gamma}) provides the relation between the choices of vacuum state in both formalisms that we were looking for
%
%
and which preserves the vacuum expectation value of the RSET.

\subsection{Degeneracy in the family of vacua \label{Sec:deg}}

We have mentioned that two different conformal Killing vectors, which in principle prescribe different vacuum states, could lead after a more careful examination to the same vacuum state. This degeneracy can be determined by studying the reparametrizations (\ref{eq:rep}) which leave the notion of vacuum unchanged. 

In the formalism of Fock quantization, these reparametrizations should give a null beta coefficient of the corresponding Bogoliubov transformation \cite{birrell-davies},
\begin{equation}
\beta(\omega,\omega')=\frac{1}{2\pi}\sqrt{\frac{\omega}{\omega'}}\int_{-\infty}^{+\infty}\text{d}u\,\exp[-i\omega u]\exp[-i\omega' U(u)]~.
\label{eq:betabog}
\end{equation}
However, it does not seem easy to find these functions $U(u)$ from the condition $\beta(\omega,\omega')=0$.

On the other hand, if two vacuum states are actually equal, the expectation values of any observable in these states must be coincident. In particular, this condition over the RSET is
\begin{equation}
\langle0_\xi|\hat{T}_{\mu\nu}|0_\xi\rangle=\langle0_{\xi'}|\hat{T}_{\mu\nu}|0_{\xi'}\rangle~,
\label{eq:cond}
\end{equation}
where $\xi$ and $\xi'$ are two conformal Killing vector fields,
\begin{align}
\xi:&=\partial_u+\partial_v~,
\label{eq:kill1}\\
\xi':&=\partial_U+\partial_V=a(u)\partial_u+b(v)\partial_v~.
\label{eq:kill2}
\end{align}
Logically, it might be possible that two different vacuum states share the same expectation value of the RSET although they differ in other observables (we will return to this point later). However, in any back-reaction analysis in semiclassical general relativity these two hypothetically different vacua will lead to the same effects.

Using the system of coordinates $(u,v)$ to evaluate the identity (\ref{eq:cond}) we get two conditions:
\begin{align}
\langle0_\xi|\hat{T}_{uu}|0_\xi\rangle&=\langle0_{\xi'}|\hat{T}_{uu}|0_{\xi'}\rangle~,\nonumber\\
\langle0_\xi|\hat{T}_{vv}|0_\xi\rangle&=\langle0_{\xi'}|\hat{T}_{vv}|0_{\xi'}\rangle~.
\label{eq:cond1}
\end{align}
The $\mu=u$, $\nu=v$  component, which encodes the conformal anomaly, does not depend on the vacuum state and therefore provides a simple identity.

Performing the calculation separately for the $u$ sector (everything applies equally to the $v$ sector) the relevant condition to analyze is
\begin{equation}
A_{uu}^\xi=A^{\xi'}_{uu}~.
\end{equation}
When using the expression for $A_{\mu\nu}^\xi$ given by $(\ref{eq:covrset})$, this equation turns out to be equivalent to an ordinary third-order differential equation for the reparametrization function $U(u)$:
\begin{equation}
\frac{\text{d}U}{\text{d}u}\frac{\text{d}^3U}{\text{d}u^3}-\frac{3}{2}\left(\frac{\text{d}^2U}{\text{d}u^2}\right)^2=0~.
\label{eq:diffu}
\end{equation}
This differential equation is well known: its three-parameter family of solutions is~(see, for instance, \cite{Fabbri:2005mw})
\begin{equation}
U(u)=\frac{au+b}{cu+d}~,
\label{eq:moebius}
\end{equation}
with $a$, $b$, $c$ and $d$ being real parameters which must verify $ ad-bc =\pm 1$. These transformations constitute the real Möbius group $\mathcal{M}_\mathbb{R}(\mathbb{C}_\infty)$ of conformal transformations of the extended complex plane $\mathbb{C}_\infty=\mathbb{C}\cup\{\infty\}$ which map the real line $\mathbb{R}$ to itself. The subgroup $\mathcal{M}_\mathbb{R}^+(\mathbb{C}_\infty)$ of solutions with $ad-bc=+1$ arises when we add the restriction of preserving the time orientation of spacetime and is  isomorphic to $\text{PSL}(2,\mathbb{R})=\text{SL}(2,\mathbb{R})/\mathbb{Z}_2$.
 
All the elements of this group can be expressed in terms of (at most four) compositions of the following fundamental transformations:
\begin{itemize}
\item  translations,
\begin{equation}
u\rightarrow U(u)=u+\alpha,\ \ \ \ \ \alpha\in\mathbb{R};
\label{eq:transl}
\end{equation}
\item dilatations,
\begin{equation}
u\rightarrow U(u)=\lambda u,\ \ \ \ \ \lambda>0
\label{eq:dil};
\end{equation}
\item inversions,
\begin{equation}
u\rightarrow U(u)=-\frac{1}{u}.
\label{eq:inv}
\end{equation}
\end{itemize}
The inequality in (\ref{eq:dil}) and the minus sign in (\ref{eq:inv}) are due to the condition of preservation of time orientation, $\text{d}U/\text{d}u>0$. Each of these transformations separately constitutes a subgroup (for the inversion we must add the identity map $U(u)=u$).

One can evaluate the beta Bogoliubov coefficients (\ref{eq:betabog}) for each of these sets of reparametrizations to determine whether they leave invariant not only the RSET, but also the vacuum state. By using the integral representation of the Dirac delta function and remembering that $\omega,\omega'\in\mathbb{R}^+$, it is straightforward to see  that  (\ref{eq:transl}) and (\ref{eq:dil}) leave the vacuum state unaffected. As for (\ref{eq:inv}), we can use complex analysis to evaluate the Cauchy principal value of the beta Bogoliubov coefficient; the result is that the vacuum state is also invariant under these transformations.



We have shown that the transformations (\ref{eq:transl}-\ref{eq:inv}) and hence the arbitrary compositions of them (\ref{eq:moebius}) leave the vacuum state unaffected and that they are, in fact, the only reparametrizations $u\rightarrow U(u)$ that do so. These results are similar to those obtained in \cite{Agullo:2006um}.

If we repeat these arguments for a general reparametrization induced by (\ref{eq:kill2}), we can see that the degeneracy on each vacuum state is dictated by the group
\begin{equation}
\text{G}=\text{PSL}(2,\mathbb{R})\times \text{PSL}(2,\mathbb{R})\label{eq:group}~.
\end{equation}
As it should be, this group includes the Poincar\'e group. Indeed it contains spacetime translations [Eq. (\ref{eq:transl}) applied to $u$ and $v$], and 
the proper orthochronous Lorentz transformations given by the dilatations $u\rightarrow\lambda u$, $v\rightarrow(1/\lambda)v$ with $\lambda>0$.

It is well known that the conformal group of any (1+1)-dimensional spacetime is infinite-dimensional. In general, for a ($1+q$)-dimensional (compactified) Minkowski spacetime with $q>1$ one can identify the conformal group as $\text{SO}(2,q+1)/\mathbb{Z}_2$. This cannot be extrapolated to the $q=1$ case: The conformal group is bigger than the group of global conformal transformations $\text{SO}(2,2)/\mathbb{Z}_2$. However, it is interesting to note that the action of $\text{SO}(2,2)/{\mathbb{Z}_2}$ decouples over null coordinates precisely as \cite{Schottenloher:2008zz}
\begin{equation}
\text{SO}(2,2)/{\mathbb{Z}_2}\simeq \text{PSL}(2,\mathbb{R})\times \text{PSL}(2,\mathbb{R})=\text{G}\label{eq:group1}.
\end{equation}
In view of this, the group of reparametrizations which dictates the degeneracy of each vacuum state is precisely $\text{SO}(2,2)/\mathbb{Z}_2$. The two copies of $\text{PSL}(2,\mathbb{R})$ correspond to the reparametrizations in the $u$ and $v$ sectors.

We can interpret this result in a different way: each complex structure (\ref{eq:Jdef}) carries a representation of $\text{SO}(2,2)/{\mathbb{Z}_2}$ which leaves it invariant. This invariance is reflected in the invariance of the corresponding vacuum state under the action of this group \cite{Corichi:2006zv}. This invariance is connected with the adjective ``conformal'' attached to the vacuum states defined by (\ref{eq:Jdef}). If we follow the convention used in the literature, this adjective seems to point out that the vacuum state is invariant under conformal transformations. However, it is well known that the trace anomaly of the RSET is due to the breaking of conformal symmetry by the vacuum state. In spite of this lack of invariance under the {\em full} group of conformal transformations (composed by global and local transformations), we have seen that they are invariant under the {\em restricted} conformal group $\text{SO}(2,2)/{\mathbb{Z}_2}$ (composed only by global transformations). Taking this restricted group as the group of symmetries at the quantum level is common in Conformal Field Theory \cite{Schottenloher:2008zz}. In fact, this symmetry of these vacuum states is related with the Hawking effect \cite{Agullo:2010hi}.

\section{\label{Sec:mott}Clearing up a subtle point}

The RSET is the starting point to investigate the effects of vacuum polarization in
gravitational dynamics. Some particularly interesting situations to look for these effects are the collapse of stellar structures towards the formation of black holes. 

For concreteness let us consider an scalar field in a (1+1)-geometry representing 
the $(t,r)$ sector of a static four-dimensional Schwarzschild black hole  (this model was first analyzed in~\cite{dfu1}):
\begin{align}
ds^2&=-\left(1-{2m \over r}\right)dt^2+ \left(1-{2m \over r}\right)^{-1}dr^2 \nonumber\\
&=-\left(1-{2m \over r}\right)dudv~,
\end{align}
where $u$ and $v$ are null coordinates defined by
\begin{align}
&u=t-r^*~, \qquad v=t+r^*~, \\
r^*=r&+{1 \over 2\kappa}\ln\left({r \over 2m}-1\right)~, \qquad\kappa={1 \over 4m}~.
\end{align}
When analyzing the RSET in this spacetime within the effective action formalism, one has to solve the field equation (\ref{eq:varphieq}). Working directly in $(u,v)$ coordinates it is easy to see that its general solution can be written as
\begin{eqnarray}
\varphi &&= \ln \left(1- {2m \over r} \right) +f(u) +g(v)~,
\label{eq:phifg}
\end{eqnarray}
where the last two terms provide the general solution of the homogeneous equation. By looking at this expression, one might be led to think that, except for rather special functions $f(u)$, the field solution $\varphi$ will be non-regular at the horizon and the RSET will be divergent there~\cite{Mottola:2006ew}. However, one can alternatively write the general solution of (\ref{eq:varphieq}) in the entirely equivalent form
\begin{align}
\varphi = \ln \left(1- {2m \over r} \right) + \kappa u + \tilde f(u) + \tilde g(v)~.
\label{eq:phitildefg}
\end{align}
Then, it is not difficult to see that the divergences of the first two terms at the future event horizon cancel each other. From this point of view, one might be led to think that only very special functions $\tilde{f}(u)$ would yield singular behaviors at the future event horizon. However, the two expressions (\ref{eq:phifg}), (\ref{eq:phitildefg}) are entirely equivalent (they are connected by a redefinition of $f(u)$) at a formal level. Therefore, without further physical ingredients it appears not possible to judge whether it would be natural or not that a selection of vacuum state leads to a divergent behavior at the horizon.

As an example of the nature of these additional ingredients, one can study the behavior of vacuum states in dynamical geometries. In~\cite{Barcelo:2007yk} it was explicitly shown within the point-splitting formalism that indeed the RSET in dynamically collapsing geometries never diverges at horizon formation (in accordance with Fulling-Sweeny-Wald's theorem~\cite{fsw}). In these calculations it is only assumed that the vacuum state at past infinity corresponds to the natural Minkowskian vacuum (more technically, one only needs to assume that the initial vacuum state is of Hadamard form). In these situations it was shown that the relevant conformal factor to use in expression~(\ref{eq:rset}) when collapsing towards the formation of a Schwarzschild black hole is
\begin{align}
C_{\rm dyn}=C_{\rm static}{du \over dU}=\left(1- {2m \over r} \right){du \over dU}~,
\label{eq:Cdynamics}
\end{align}
where $U$ and $u$ represent, respectively, the past and the future outgoing null coordinates (the $v$ sectors can be neglected in what concerns any potential divergence at the horizon). It is well known that if a stationary horizon forms in the collapse the asymptotic form of the relation $U(u)$ is \cite{hawking-radiation}
\begin{align}
U \simeq U_H -Ae^{-\kappa u}~,
\label{eq:asymprel}
\end{align}
where $U_H,A$ are suitable constants. Now, by using the translation map (\ref{eq:relation1}) and the previous expression, it is easy to conclude that at the horizon
\begin{align}
\varphi &= \ln C_{\rm dyn} = \ln C_{\rm static} + \ln {du \over dU} 
\nonumber \\
&\simeq \ln \left(1- {2m \over r} \right) + \kappa u -\ln(\kappa A)~.
\label{eq:phitildyn}
\end{align}
Therefore, apart from the first term in (\ref{eq:phitildyn}), which is present even in static configurations, the dynamics of the collapse always generates a term in the form $+\kappa u$ which precisely cancels the potential divergence in the RSET~\cite{Barcelo:2007yk} (after collapse the system always sets in an Unruh-like state~\cite{unruh-vacuum}). To this last expression one could add transient particle fluxes through the geometry in the form of regular functions $\tilde f,\tilde g$, matching then expression (\ref{eq:phitildefg}), without affecting the main result: The RSET in any collapsing geometry producing a horizon is perfectly regular at horizon formation. 

To end this brief section, let us comment that although the RSET will not diverge in the eventual case of horizon formation, this does not mean that it cannot become very large before the actual formation of the horizon. In principle, it could become so large in some situations that its semiclassical back-reaction in the geometry might completely modify the final fate of collapse avoiding horizon formation \cite{Barcelo:2007yk,visser-sdh,mottola-lectures}. Under which circumstances this could happen is a matter of active investigation.

\section{\label{Sec:summ}Summary and conclusions}

In this paper we have explicitly compared the structural form of the renormalized stress-energy tensor (RSET) of a massless scalar field over a (1+1) curved spacetime as obtained by two different  strategies: normal-mode construction of the field operator and one-loop effective action. The problem in comparing these two expressions is that they use different fiduciary structures. Once a translation dictionary has been established,  we have shown that these two structures
are actually equal.

We have put special emphasis in explaining how and where  the information associated with the selection of different vacuum states in both formalisms is encoded. We have also provided a translation map that, given a vacuum state in one formalism, tells us how to select the same vacuum state when working in the other formalism.

The overall aim of this  analysis is to facilitate the comparison between specific calculations and results obtained by using these two apparently different approaches.
In particular, we have used our translation dictionary to clear up a potentially misleading way to look at the problem of how gravitational collapse is affected by semiclassical corrections.

\acknowledgments

This work originated from a conversation we had with Emil Mottola. We thank him for that and for all his comments during its development. We also want to thank V\'ictor Aldaya, Manuel Calixto, Julio Guerrero and Guillermo A. Mena Marug\'an for helpful comments and discussions. Financial support was provided by the Spanish MICINN through the projects FIS2008-06078-C03-01 and FIS2008-06078-C03-03 and by the Junta de Andaluc\'{\i}a through the project FQM219. R.C. acknowledge support from CSIC through a JAE-predoc fellowship.

\end{document}